\documentclass[]{spie}  

\newif\ifpdf
\ifx\pdfoutput\undefined
\pdffalse 
\else
\pdfoutput=1 
\pdftrue \fi
\ifpdf
\usepackage[pdftex]{graphicx}
\usepackage[pdftex]{color}
\DeclareGraphicsExtensions{.jpg,.png,.pdf} \else
\usepackage[dvips]{graphicx}
\usepackage[dvips]{color}
\DeclareGraphicsExtensions{.eps} \fi

\title{Slow and stored light and optical depth}

\author{ 
  M.\ Klein\supit{a,b}, 
  Y.\ Xiao\supit{a}, 
  A.\ V.\ Gorshkov\supit{b},
  M.\ Hohensee\supit{a,b},
  C.\ D.\ Leung\supit{a,b}, 
  M.\ R.\ Browning\supit{a,b}, 
  D.\ F.\ Phillips\supit{a},
  I.\ Novikova\supit{c},
 and R.\ L.\ Walsworth\supit{a,b} 
 \skiplinehalf
    \supit{a}Harvard-Smithsonian Center for Astrophysics, Cambridge, MA,
  02138 USA \\
  \supit{b}Department of Physics, Harvard University, Cambridge, MA, 02138
  USA \\
  \supit{c}Department of Physics, College of William \& Mary,
  Williamsburg,
  VA 23185, USA
}

\date{\today}
\begin{document}
  \maketitle


\begin{abstract}

  We present a preliminary experimental study of the dependence on optical depth of slow and
  stored light pulses in Rb vapor. In particular, we characterize the efficiency of slow and stored light as a
  function of Rb density; pulse duration, delay and storage time; and
  control field intensity. Experimental results are in good
  qualitative agreement with theoretical calculations based on a
  simplified three-level model at moderate densities.

\end{abstract}


\maketitle

\section{Introduction}\label{intro}

Applications of slow and stored light based
on electromagnetically-induced transparency (EIT), in both quantum information
processing and optical communication, will benefit from improved 
efficiency and delay-bandwidth product.
In quantum information, stored light has emerged as a promising
technique for applications such as single-photon generation on
demand~\cite{kimbleQM,kuzmichQM,lukinQM,kuzmichQM2,chenQM} and quantum
memories~\cite{chouQM,chaneliereQM,eisamanQM} and
repeaters~\cite{DLCZ,chaneliereQM2,felintoQM}. However, practical
applications will require significant improvements in the efficiency of
writing, storing, and retrieving an input photon state beyond values
achieved to date~\cite{novikovaEff,felintoEff,lauratEff,vuleticEff}.
In classical communications, optical buffers require adjustable delay
time (i.e., group index) for input signal pulses with minimal pulse distortion and loss~\cite{tuckerSL} and
high compression of the input pulse for high data density inside the
delay medium. However, high efficiency and large
delay-bandwidth product have not fully been realized in all-optical systems.
Here we present a preliminary study of the optimization of slow and
stored light using EIT in warm Rb vapor as a function of optical depth, laser power, and input pulse bandwidth.

In EIT, a weak signal pulse propagates with very slow group
velocity through an otherwise absorbing medium~\cite{hau99}, which is established
by a strong control field.
The reduced group velocity with which a signal propagates under EIT is
the result of the transfer of photonic excitation in the pulse to the
collective spin coherence of the atoms.  The group velocity is
proportional to the control field intensity, and can be reduced to
zero by shutting off the control light.  In this case the photonic
information of the pulse is completely transferred to the atoms.  This
coherent transfer is reversible, and thus after some storage time the
pulse may be retrieved, converting it back into photonic excitation
when the control light is turned back on.

Optimization of slow and stored light requires reducing the
absorption of signal light and spin decoherence, as well as increasing optical depth to achieve a large delay-bandwidth product~\cite{lukin03rmp}.
In warm atom vapor cells, signal light absorption is typically driven
by incomplete polarization of the atomic medium due to such mechanisms
as radiation trapping and competing nonlinear processes.
Atomic decoherence arises from mechanisms such as
collisions with buffer gas atoms and cell walls, diffusion 
out of the laser beam, and residual magnetic field gradients.

Modeling the atomic medium as a collection of stationary three-level
$\Lambda$-systems (Fig.~\ref{LevelsAndSetup}a), we find that 
slow and stored light properties improve with increasing rubidium
density or optical depth.  The EIT bandwidth, $\gamma_{EIT}$, for
moderate to large optical depths is found to be~\cite{hauNL,fleischhauerDarkState}
\begin{equation}
 \gamma_{EIT} = \frac{|\Omega_{C}|^{2}}{\sqrt{\gamma g^{2}NL/c}}
\label{e.EITwidth}
\end{equation}
where $\Omega_C$ is the Rabi frequency of the control field,
$\gamma$ the excited-state line width, $N$ the total number of
three-level atoms, $L$ the length of the medium (the optical depth,
$d=g^{2}NL/\gamma c$), $g$ the light-atom coupling
coefficient, and $c$ the speed
of light.
Similarly, the group velocity $v_{g}$ resulting from the steep
resonant dispersion for the signal pulse in the slow light medium
is~\cite{hauNL,fleischhauerDarkState}
\begin{equation}
 v_{g} = \frac{c}{1+g^{2}N/|\Omega_{C}|^{2}} \approx \frac{c|\Omega_{C}|^{2}}{g^{2}N}.
\end{equation}
The absolute pulse delay, $\Delta T_{\mathrm{abs}}$, is then given by the length of
the medium divided by the group velocity and we find that
$\Delta T_{\mathrm{abs}}\propto d$, that is the delay is proportional to the
optical depth. 
%
%
A careful theoretical study of storage efficiency $\eta$ (defined as
the ratio of the number of output photons to the number of input
photons), optimized for a
given optical depth~\cite{gorshkovOpt}, finds that the optimal temporal pulse width $T_{opt}\sim d$, and finds that $\eta$ scales as $1-19/d$ at high optical depth, and increases more rapidly at lower $d$.  This treatment implies continued efficiency improvement with increasing optical depth.  
%
In practice, experiments have found that slow and stored light efficiency improves with
increasing optical depth for modest optical depths. However, at larger
optical depths, slow and stored light
efficiency reach a maximum and then degrade.  Note that a procedure for optimizing
stored light efficiency for a given $d$, using the time-reversed
output pulse to determine the shape of the input pulse, has been
developed theoretically~\cite{gorshkovOpt} and directly
implemented experimentally~\cite{novikovaOpt}.

In the present study we report preliminary
experimental investigations of the behavior of slow and stored light
over a wide range of optical depths, by varying the temperature of Rb
in a vapor cell.  Measuring coherence loss rates and including them in
the efficiency optimization model yields good agreement between theory
and experiment as $\eta$ falls off with increasing atomic density.  We
focus on the role played by optical depth in coherence loss, in
particular the effect of radiation trapping, where incoherent photons
are repeatedly re-absorbed by the medium before exiting~\cite{RadTrap}.  We also report promising preliminary efficiency
results for a new cell geometry designed to reduce coherence loss due
to radiation trapping.

\section{Experimental setup}\label{ExpSetup}

\begin{figure}
  \centering{\includegraphics[width=0.6\columnwidth]{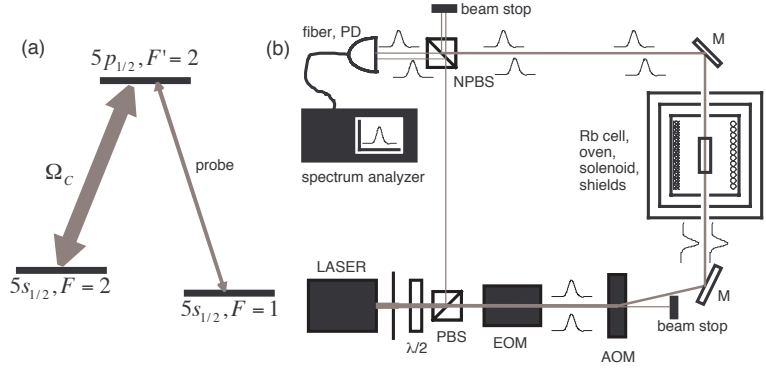}}
  \caption{ (a) The relevant ${}^{87}$Rb atomic energy levels in a 
    $\Lambda$-scheme used for EIT, slow,
    and stored light measurements.  (b) Schematic of the
    apparatus. Output of laser is split with a half-wave plate
    ($\lambda/2$) and polarizing beam  splitter (PBS) into a reference
    beam and main beam; electro-optic (EOM) and acousto-optic (AOM)  modulators
    shape the signal and control fields; the vapor cell is housed
    inside an oven, solenoid, and magnetic
    shielding; a fast photodetector measures the beat
    between the unmodulated reference beam and the output from the
    vapor cell. \emph{See text for details.}
  }
 \label{LevelsAndSetup}
\end{figure}

We measured EIT and slow and stored light, using the $D_{1}$
transition in a warm ${}^{87}$Rb vapor.  The relevant energy levels
and coupling fields are shown in Fig.\ \ref{LevelsAndSetup}a: a strong
control field is resonant with the $F=2\rightarrow F^{\prime}=2$
transition and a weak signal field is resonant with $F=1\rightarrow
F^{\prime}=2$.
%
The optical fields were generated by an external cavity diode laser
tuned to $\lambda=795$~nm, amplified by a tapered amplifier, and
spatially filtered by a pinhole (Fig.~\ref{LevelsAndSetup}b). An
electro-optic modulator (EOM) phase modulated the laser at the ground
state hyperfine splitting ($\sim6.835$~GHz); the $+1$ sideband acted
as the signal field, with a maximum signal to control intensity ratio of
$2.5\%$; the sideband amplitude was varied to shape the desired signal pulse.  The
frequency could be varied to change the two-photon detuning $\delta$
of the EIT transition for measuring its line shape.
The overall intensity was regulated with an acousto-optic modulator (AOM),
which also shifted the frequency by $+80$~MHz.  A quarter-wave plate
($\lambda/4$) before the cell converted the beam to circular
polarization.  The collimated beam entering the cell had a diameter of
approximately $7$~mm.  The output fields were measured by sending the
laser into an optical fiber, then into a fast photodetector (PD) along with a reference beam
picked off prior to the EOM and AOM.  The control field, signal field,
or the unused $-1$ sideband could thus be measured by beat note
detection with a spectrum analyzer, at $80$~MHz,
$6.835$~GHz~$+~80$~MHz, and $6.835$~GHz$~-80$~MHz respectively.  The transmission of the
$-1$ sideband (off EIT resonance), which is $\sim6$~GHz away from any Rb resonance, was used
as a reference for determining slow light delay or storage efficiency,
where $\Delta T_{abs}$ was the difference between pulse peak times and
efficiency $\eta=$~(signal pulse area)/($-1$ sideband area).

The cell was housed in a plastic oven, which was heated by blown warm
air.  The cell temperature was varied between $40$ and $80~^{\circ}$C,
or atomic number density between $4\times10^{10}$ and
$1\times10^{12}$ cm${}^{-3}$.  Three layers of cylindrical
high-permeability shielding surrounded the oven to screen out stray
laboratory magnetic fields, and a solenoid inside was used to cancel
any small constant background field.
Three Rb vapor cells were used in our experiments; all three contained
isotopically enriched ${}^{87}$Rb and were filled with buffer gas to
confine atoms and extend their coherence life times.  The cell used
for all measurements except where specifically noted had length
$L=7.5$~cm and diameter $D=2.5$~cm, and was filled with $40$~Torr of
Ne.  


\section{Slow and stored light efficiency measurements}

We studied  slow and stored light as a function of optical depth by
measuring the delay-bandwidth product and efficiency of ``optimized''
pulses as the vapor cell temperature was varied. 
Such pulses were constructed through an iterative
optimization procedure in which a signal pulse was stored and retrieved,
its output shape measured and a new pulse stored with the shape of the
output pulse. This process is repeated until convergence. This
procedure has been shown both theoretically~\cite{gorshkovOpt} and
experimentally~\cite{novikovaOpt} to produce slow-light pulse shapes
that optimize storage efficiency for the given optical depth and laser
power.  In effect this procedure selects a pulse bandwidth that balances
 fractional delay and absorption to yield the largest output signal
pulse.

\begin{figure}
 \includegraphics[width=0.95\columnwidth]{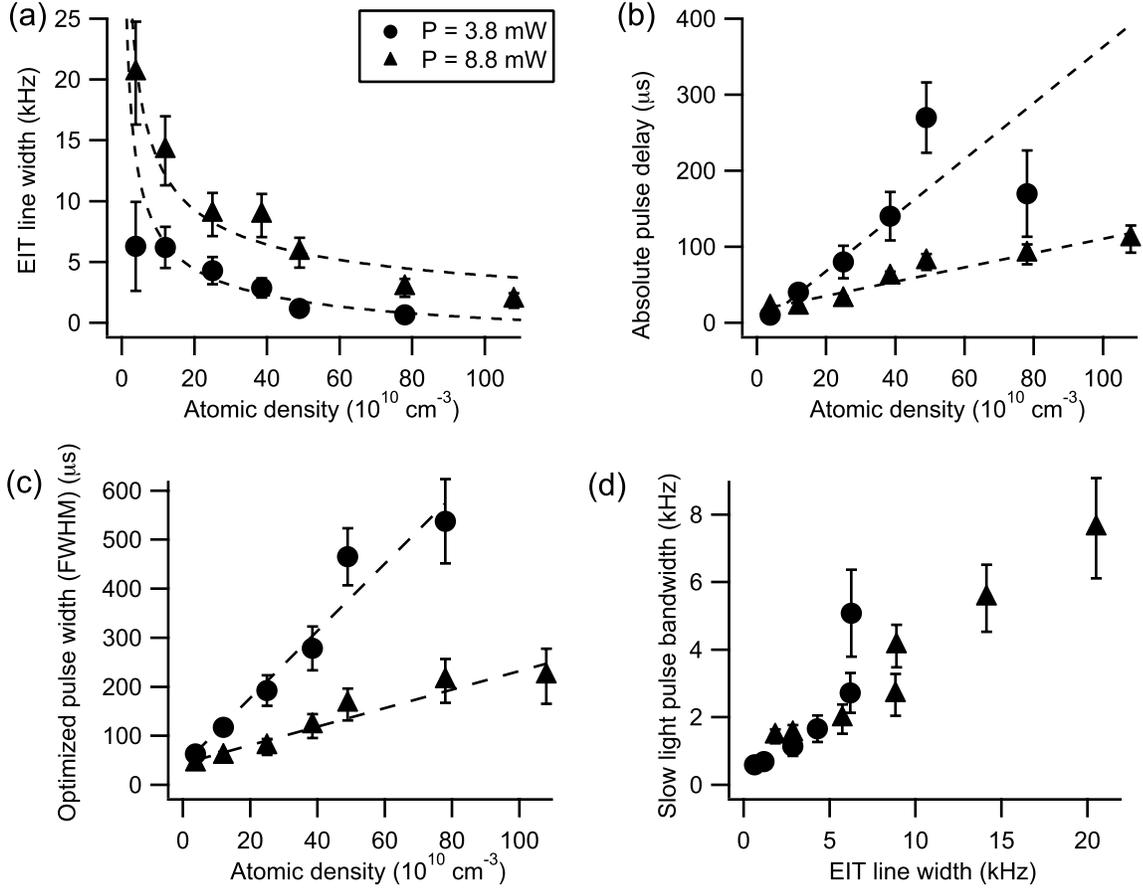}
 \caption{Measured EIT line width, slow light delay, and optimized pulse width at various
   optical depths and for two control field powers (3.8 mW and 8.8 mW).  (a) EIT line widths are consistent with power broadening and density narrowing.  (b)
   Absolute pulse delay (= time elapsed between the peaks of the
   reference pulse and the slowed pulse) and
   (c) optimized temporal pulse widths are approximately linear in atomic density. (d)
   Optimized slow light bandwidth vs.\ EIT line width.  Error bars for all measurements are
   derived from a systematic uncertainty in the Rb temperature of
   $\sim 2~^{\circ}$C and uncertainty in the laser frequency of $\sim
   150$~MHz.  Dashed lines are fits of data to the simplified three-level model described in text.}
 \label{EITandDelays}
\end{figure}

Measured EIT bandwidths and slow light delays scaled as expected with optical
depth. Figure \ref{EITandDelays}a shows EIT linewidth (FWHM, as
extracted from a Lorentzian fit) vs. atomic density, which follows the
$\gamma_{EIT} \propto 1/\sqrt{d}$ trending expected from
Eq.~\ref{e.EITwidth}. Figure~\ref{EITandDelays}b shows that absolute pulse delay
 $\Delta T_{abs} \propto d$; and Figure~\ref{EITandDelays}c shows that the optimal pulse width $T_{opt}\propto d$.  
We measured line widths and delays at two control field powers,
$3.8$~mW and $8.8$~mW, corresponding to respective intensities of $10$
and $23$~mW/cm${}^{2}$ and Rabi frequencies of $6.7$ and $10$~MHz.
Slow light delays
in Figs.\ \ref{EITandDelays}b and \ref{EITandDelays}c are
tailored for optimized stored light, obtaining neither the maximum
absolute delay (achieved for a temporally long pulse experiencing only
the steepest part of the dispersion, but having a small fraction of the
pulse in the medium at one time) nor the maximum fractional
delay (achieved for a temporally short pulse which suffers large absorption
due to frequency components outside the EIT bandwidth). 
Comparing the optimized pulse bandwidth (where $T_{opt}$ is the
temporal pulse width) to the measured EIT bandwidth
(Fig.~\ref{EITandDelays}d), we find that $1/T \sim \gamma_{EIT} / 3$,
over a large range of fractional delays (from $0.2$ to $>1.0$), a
somewhat greater pulse bandwidth than expected
theoretically~\cite{fleischhauerDarkState,gorOptimize}, where
$\gamma_{EIT}T>>1$ or even $>>10$.  Note that based on the scalings of
$\Delta T_{abs}\sim d$ and $T_{opt}\sim d$, we would expect the
optimal frequency bandwidth to be quadratic in the EIT line width
which is not clearly observed in Fig.~\ref{EITandDelays}d.

\begin{figure}
 \centering{\includegraphics[width=0.95\columnwidth]{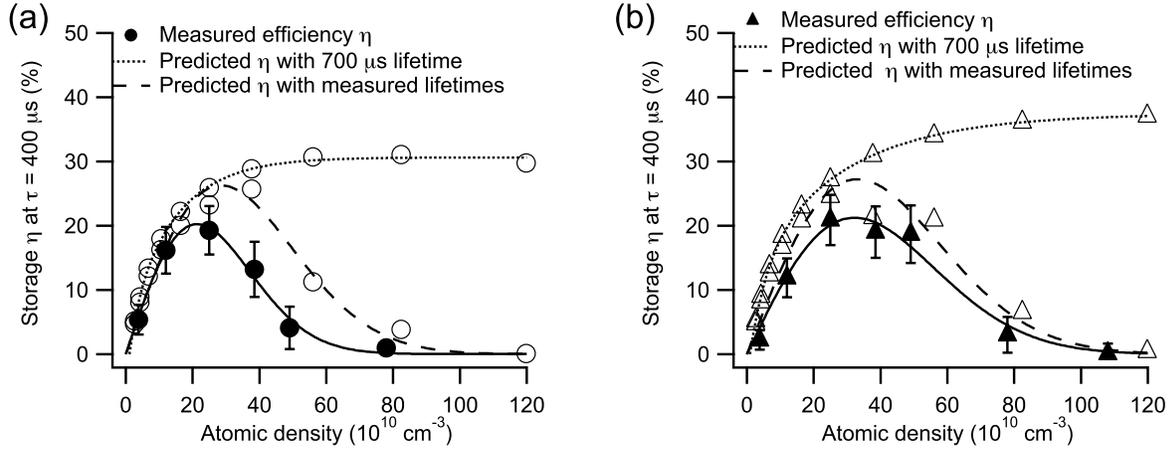}} 
 \caption{Measured and calculated storage efficiency at fixed storage
   time $\tau=400~\mu$s vs.\ atomic density for control field laser
   powers of (a) $3.8$~mW and (b) $8.8$~mW. Theoretical predictions are shown both
   for a coherence lifetime of $700~\mu$s with no density-dependent decoherence (dotted curves) and for
   the measured coherence lifetimes shown in Fig.\ \ref{oneOverEtimes}
   (dashed curves). Comparison of the two calculated results shows
   that density or temperature-dependent decoherence during the
   storage interval dominates high-density losses.  The
   remaining discrepancy between measurement and calculation is likely due to radiation trapping.  Error
   bars are from systematic uncertainty in Rb temperature of $\sim
   2~^{\circ}$C and of the laser frequency of $\sim 150$~MHz.
   \emph{Curves added to guide the eye.}  }
 \label{TheoryExp}
\end{figure}

\begin{figure}
 \centering{\includegraphics[width=0.62\columnwidth]{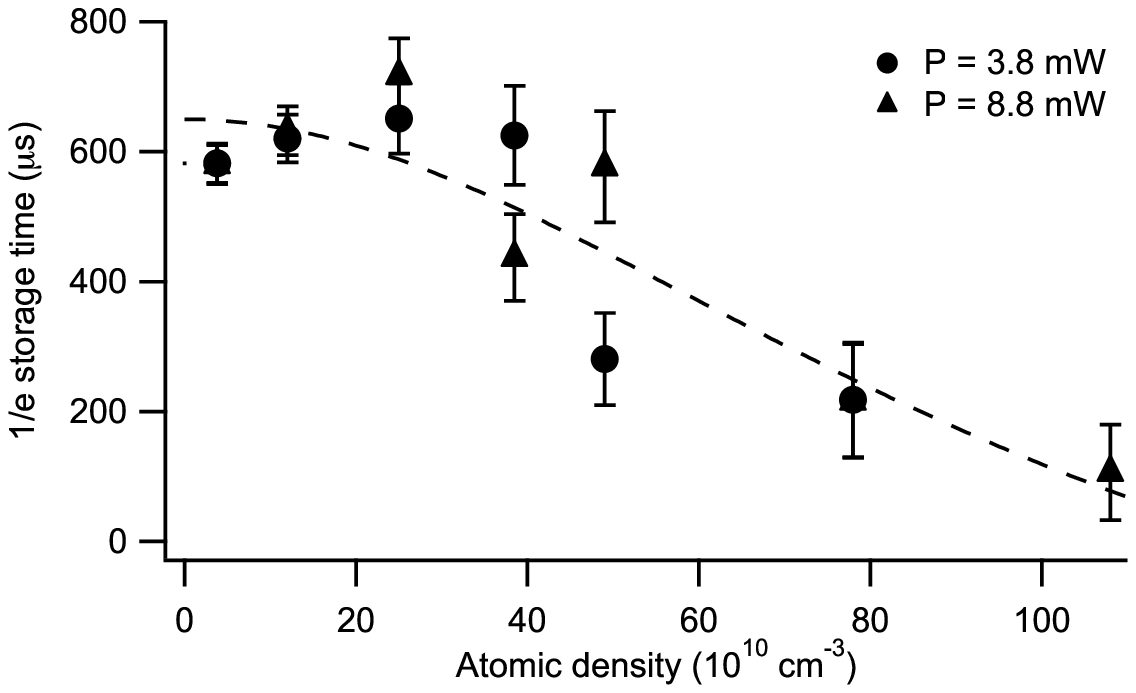}}
 \caption{Lifetimes of stored coherence as a function of atomic density, derived from measurements described in text.
   The coherence lifetime drops at high atomic density, likely due to
   the effect of temperature-dependent losses such as Rb-Rb spin
   exchange. Error bars are derived from Rb temperature and laser
   frequency uncertainty as well as statistical uncertainty in fits to
   coherence lifetimes. \emph{Dashed line added to guide the eye.}}
 \label{oneOverEtimes}
\end{figure}

Stored light efficiency
was measured for a range of rubidium densities between $4 \times
10^{10}$ cm${}^{-3}$ ($40~^{\circ}$C) and $1 \times 10^{12}$
cm${}^{-3}$ ($80~^{\circ}$C) and storage intervals from $\tau=50~\mu$s
to $\tau=1.5$~ms and plotted for $\tau=400~\mu$s in
Fig.~\ref{TheoryExp}.  The efficiencies peak at temperatures between
$60$ and $65~^{\circ}$C, falling at higher optical depth.  The $1/e$ coherence lifetimes
during storage (Fig.~\ref{oneOverEtimes}) also fall at high density.
This points to possible high-density or high-temperature phenomena
adversely affecting slow and stored light.
Increased loss during storage and readout at high optical depth agrees with
predictions from theoretical simulations when the density
dependence of decoherence is included (Fig.~\ref{TheoryExp}).  The
simulations, described fully in earlier
work~\cite{gorshkovOpt,novikovaOpt,lukin03rmp}, consist of iterated
solutions to the stored light dynamics equations for an ensemble
$\Lambda$-system.  The dashed curve in Fig.\ \ref{TheoryExp} uses the
measured rates of decoherence during storage and readout taken from Fig.\ \ref{oneOverEtimes} as
an input parameter for the spin-wave decay rate, and most closely
matches the experimental results.

\begin{figure}
  \centering{\includegraphics[width=0.9\columnwidth]{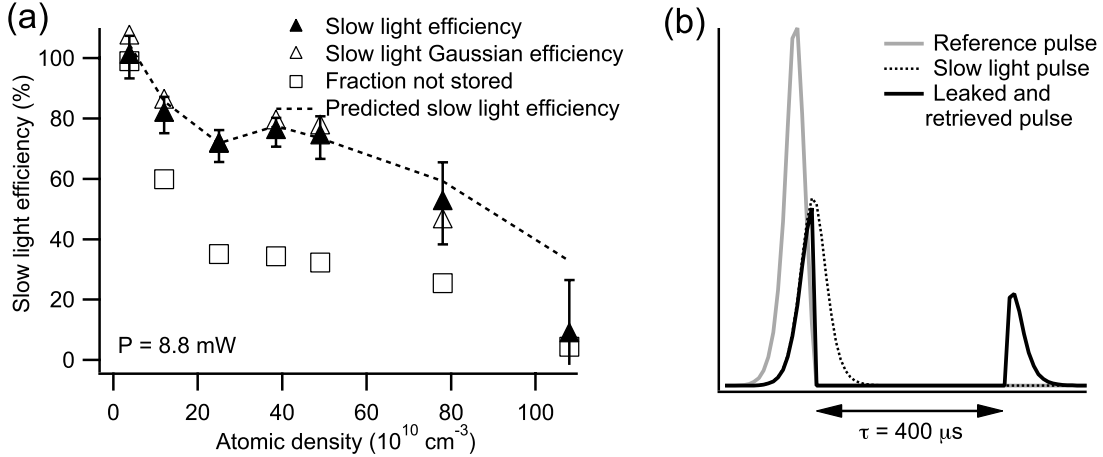}}
 \caption{ (a) Slow light efficiency (output pulse area divided by input pulse
   area) as a function of atomic density, for input pulse shapes optimized
   for maximum stored light efficiency (solid triangles) and for input Gaussian pulses with the
   same FWHM (hollow triangles).  Also shown are measurements relevant to stored light efficiency: the fraction of the optimized input pulse that
   escapes the medium before the pulse is stored (open
   squares). The dashed line combines light storage data at
   $\tau=400~\mu$s and coherence lifetime measurements to predict
   slow light efficiency (\emph{see text}) in good agreement with
   measured efficiency. Error bars are from systematic uncertainty in
   Rb temperature of $\sim 2~^{\circ}$C and in the laser frequency of
   $\sim 150$~MHz. Error bars on other data sets are of equal
   magnitude. (b) Sample data for density of approx. $8 \times
10^{11}$ cm${}^{-3}$ showing the retrieved and
   unstored pulses (dark, solid line), the slow light pulse with
   optimized shape (dotted line), and the reference pulse (gray solid
   line).  }
\label{accounting}
\end{figure}

In Figure~\ref{accounting} we verify that the loss mechanisms for
stored light are all self-consistent.  A slow light pulse that is not
trapped in the medium (the control field is left on) has loss from
imperfect EIT transmission and from pulse bandwidth lying outside the
region of maximum transmission (after normalization to a reference
pulse, this is the ``slow-light efficiency'').  A stored and retrieved
pulse will experience that same loss, and additional loss from two
sources: (i) part of the pulse escapes the medium before the pulse is
stored (fractional delay less than unity implies significant losses
from this mechanism); and (ii) atomic coherence decay during the
storage and readout periods.  Figure \ref{accounting}a shows the
directly measured slow light efficiency $\eta_{slow}$, the area ratio
of the output pulse to the reference pulse, and an inferred
$\eta_{slow,predict}$ from the storage measurement.  Using the
measured decoherence rate from Fig.\ \ref{oneOverEtimes} and the
storage readout areas at $\tau=400~\mu$s, we can back out the storage
readout at $\tau=0$; combining this with the area of the not-stored
pulse leakage we infer $\eta_{slow}$ (see Fig.~\ref{accounting}b):
\begin{equation}
\eta_{slow,predict}=\eta_{leakage}+\eta \times \exp(400~\mu s/\tau_{coherence}),
\end{equation}
where $\eta_{leakage}$ is the leaked pulse area divided by the
reference pulse area, and $\tau_{coherence}$ is the atomic coherence
lifetime from Fig.\ \ref{oneOverEtimes}.  The measured and inferred
slow light efficiencies match very well, only deviating significantly
at the highest atomic density, where the term with
exp$(400~\mu$s$/\tau_{coherence})$ becomes very sensitive to the short
coherence lifetime measurement, which is more error prone due to the
low SNR in the signal transmission.

As indicated from both the storage efficiency and the comparison
between slow and stored light, at high Rb densities the effectiveness
of the medium is reduced. We believe that several mechanisms are
responsible for this. First, at higher densities, temperature or
density-dependent losses such as Rb-Rb spin exchange
become more significant, contributing to decoherence even during the
storage interval (see Fig.~\ref{oneOverEtimes}). Second, in the
presence of light at high Rb densities, radiation trapping becomes
significant: light absorbed and then incoherently radiated by
one rubidium atom is reabsorbed by another rubidium atom, and this process
leads to decoherence and loss of efficiency. In the next sections, we
present preliminary investigations of high-density losses.

\section{Optical depth and radiation trapping measurements} 

Direct measurements of large optical depths are difficult due to the
large corresponding absorption of the optical field. To characterize
the optical depth for comparison with the optimization models
described above, we used optical absorption line width as a proxy for
the optical depth. The signal field was detuned from two-photon
resonance by changing the frequency driving the EOM and the signal
intensity monitored as the one-photon (laser) detuning was swept
through resonance. The line width of the signal field absorption serves as a proxy for the optical depth because the logarithm of the measured absorption is proportional to the product of the detuning and
the optical depth.

\begin{figure}
\centering{\includegraphics[width=0.80\columnwidth]{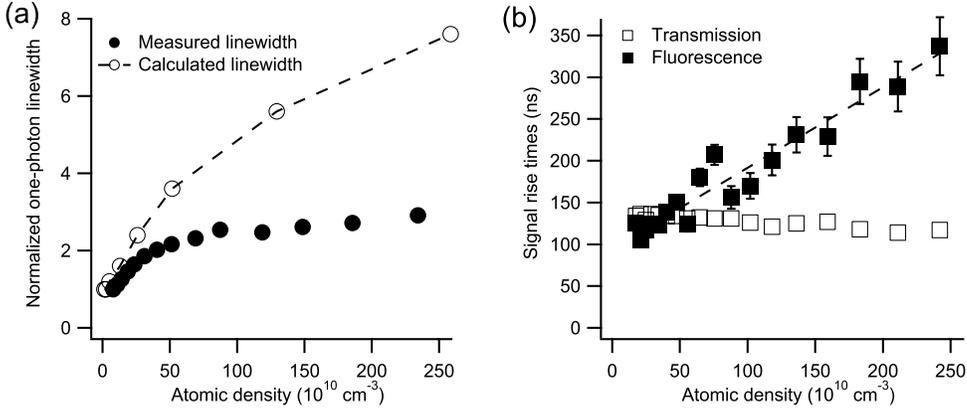}}
\caption{ (a) Measured and calculated optical line widths for the signal
  field when detuned from two-photon resonance as a function of Rb
  density, consistent with an effectively reduced optical depth at high densities due to radiation trapping.  (b) Fluorescence signal rise time
  increases as the atomic density is increased, whereas transmission signal rise time has minimal density dependence, indicating that
  absorbed and re-emitted photons are trapped in the medium for a
  longer time due to the large transverse optical depth. (\emph{See
    text.}) Error
  bars are smaller than data points except where shown (derived from uncertainty in Rb
  temperature).}
\label{RadTrap}
\end{figure}

Figure~\ref{RadTrap}a shows a comparison of the measured one-photon linewidth and
the expected linewidth from a simple numerical model based on an optically thick
sample of three-level $\Lambda$-systems. At high densities (and optical depths), the measured
linewidth stops growing while the model shows continued line
broadening. We attribute this to radiation trapping: photons scattered
inside the medium depolarize the rubidium ground state.  Fewer atoms are polarized in the signal field channel, leading to bothdecoherence and absorption of the signal field.

We also directly measured the presence of radiation trapping in our
vapor cells, with a simplified apparatus.
Light from an extended cavity diode laser was focused through an AOM
with a small spot size to allow for rapid rise times in the AOM.  This
AOM output was turned on and off at a rate of $1$~MHz with a rise time
of $\sim 75$ ns. The beam diameter was then expanded to approximately
$2$~mm and sent into a Rb vapor cell housed inside a small plastic
oven, unshielded from stray laboratory magnetic fields, and warmed
with resistive heaters to temperatures between $54$ and
$90~^{\circ}$C.  Output light from the cell was measured with two
photodetectors: one for transmission, and the other for fluorescence
mounted to the side of the cell oven where a lens close to the cell
focused scattered light onto the detector.  When light was turned on
via AOM modulation, the rise time of the signal was observed in both
detectors.

Evidence for radiation trapping is shown in Fig.\ \ref{RadTrap}b.  At
high atomic densities, the rise time in the fluorescence detection
signal increases significantly.  The greater the
transverse optical depth, the greater the number of absorption and
re-emission cycles that occur before absorbed photons exit the medium, so the
rise time seen in fluorescence detection should grow with the
transverse optical depth.  In Fig.\ \ref{RadTrap}b we see rise times
up to $350$~ns, whereas the lifetime for the excited state
of the $5p_{1/2}$ state of Rb is $\sim 28$~ns, indicating many scattering events for each absorbed
photon.

\section{New vapor cell geometry}

In order to minimize the decoherence resulting from radiation trapping
in our vapor cells, we recently developed a cell designed to minimize its effects.
A 15 cm long, 1.2 cm diameter cell with N${}_{2}$ buffer gas was designed to reduce
radiation trapping in three ways: (i) the aspect ratio is four times
larger than previous cells, allowing fluorescence to escape the cell
in the transverse direction with fewer depolarizing interactions with
Rb atoms; (ii) the absolute length of the cell is twice that of
previous vapor cells allowing the equivalent optical depth to be
reached at lower Rb density, minimizing density-dependent effects,
including spin exchange; (iii) N${}_{2}$ buffer gas acts to quench
radiation trapping by collisionally de-exciting Rb atoms before the
atoms fluoresce, preventing the emission of unpolarized photons that
could destroy the coherence of nearby atoms.

Initial results of stored light efficiency in this cell appear
promising.  For a laser power of $P=4.5$~mW and a temperature of
$58~^{\circ}$C (the same longitudinal optical depth as the density $=40\times
10^{10}$~cm${}^{-3}$ data in Fig.\ \ref{TheoryExp}), efficiency of
$\eta=40\%$ was achieved.  This is three times greater than the $\eta$
for $P=3.8$~mW and twice as great as for $P=8.8$~mW at the same
optical depth.  Future measurements will further explore this promising avenue towards high-efficiency light storage in Rb vapor cells.


\section{Conclusions}

We reported a preliminary experimental study of slow and stored light
at a variety of Rb densities (i.e., optical depths), and as a function of pulse duration, delay and storage time, and
  control field intensity. Experimental results are in good
  qualitative agreement with theoretical calculations based on a
  simplified three-level model at moderate densities; and indicate that radiation trapping is an important limitation at high atomic density. We have developed a new cell geometry which we expect to reduce losses and improve efficiency
through reduced radiation trapping and density-dependent
decoherence.

We are grateful to A.\ Glenday for useful discussions. This work was
supported by ONR, DARPA, NSF, and the Smithsonian Institution.

\end{document}